\documentclass[%
aps,
pra,
longbibliography,
 amsmath,amssymb,
 reprint,%
]{revtex4-1}

\usepackage{graphicx}% Include figure files
\usepackage{dcolumn}% Align table columns on decimal point
\usepackage{bm}% bold math
\usepackage[utf8]{inputenc}
\usepackage[T1]{fontenc}
\usepackage{mathptmx}
\usepackage{siunitx}

\usepackage[compact]{titlesec}
\titlespacing{\section}{0pt}{10pt}{3pt}

\usepackage{xcolor}
\usepackage[normalem]{ulem}

\begin{document}
\setlength{\belowcaptionskip}{-10pt}
\raggedbottom
\parskip=5pt
\preprint{AIP/123-QED}

\title{Direct observation and evolution of electronic coupling between organic semiconductors}%

\author{Sameer Vajjala \surname{Kesava}}
\email[Correspondence: ]{sameer.vajjalakesava@physics.ox.ac.uk,}
\affiliation{Department of Physics, University of Oxford, OX1 3PU, England, UK}
\author{Moritz K. \surname{Riede}}
\email[]{moritz.riede@physics.ox.ac.uk} 
\affiliation{Department of Physics, University of Oxford, OX1 3PU, England, UK}

%\date{\today}%

\begin{abstract}
The electronic wavefunctions of an atom or molecule are affected by its interactions with its environment. These interactions dictate  electronic and optical processes at interfaces, and is especially relevant in the case of thin film optoelectronic devices such as organic solar cells. In these devices, charge transport and interfaces between multiple layers occur along the thickness or vertical direction, and thus such electronic interactions between different molecules -- same or different -- are crucial in determining the device properties. Here, we introduce a new in-situ spectroscopic ellipsometry data analysis method called DART with the ability to directly probe electronic coupling due to intermolecular interactions along the thickness direction using vacuum-deposited organic semiconductor thin films as a model system. The analysis, which does not require any model fitting, reveals direct observations of electronic coupling between frontier orbitals under optical excitations leading to delocalization of the corresponding electronic wavefunctions with thickness or, equivalently, number of molecules away from the interface in C60 and MeO-TPD deposited on an insulating substrate (SiO$_2$). Applying the same methodology for C60 deposited on phthalocyanine thin films, the analyses shows strong, anomalous features -- in comparison to C60  deposited on SiO$_2$ -- of the electronic wavefunctions corresponding to specific excitation energies in C60 and phthalocyanines. Translation of such interactions in terms of dielectric constants reveals plasmonic type resonance absorptions resulting from oscillations of the excited state wavefunctions between the two materials across the interface. Finally, reproducibility, angstrom-level sensitivity and simplicity of the method are highlighted showcasing its applicability for studying electronic coupling between any vapor-deposited material systems where real-time measurements during thin film growth are possible.  
\end{abstract}

%Use showkeys class option if keyword
                              %display desired
\maketitle
%\tableofcontents

\section{INTRODUCTION}
The electronic wavefunctions of an atom in a solid depend on the positions of the electrons and the nuclei. Changes in the energy levels occur due to the overlap of the ground state electronic wavefunctions primarily of the frontier orbitals between neighboring atoms or molecules, and determines the optoelectronic properties of a solid, e.g. leading to the emergence of bands in semiconductors \cite{AshcroftNeilW1976Ssp,SimonSteven}. Understanding how the energy levels in an atom evolve from isolation to a many-atom solid is one of the most important fundamentals of solid state physics required for explaining the properties of solids such as metals, semiconductors and insulators.
 
In the case of organic semiconductors, comprised typically of large molecules, understanding the effects of intermolecular interactions on the electronic wavefunctions is crucial for understanding the physics at interfaces since the interfacial phenomena, e.g. transfer of a charge between a donor and an acceptor in solar cells, are affected by the local environment \cite{Azzouzi2018_PRX, Eisner2019_JACS, Athanasopoulos2019_JPCL, Kahle2018_JPCC, Menke2018_JOule, Gelinas2014_Science} to which the frontier orbitals react. This is relevant not just for organic semiconductors but any electronic material such as conductive oxides, chalcogenides and metals  forming the layers of optoelectronic thin film devices where interfacial processes play a critical role in the device properties, and especially more in the case of vertical devices (where charge transport occurs along the thickness direction in contrast to thin film field-effect transistors). 

In-situ optical spectroscopic techniques such as in-situ differential reflectance spectroscopy (DRS) \cite{Forker2012_APSC, Gruenewald2016_PRB} and reflection anisotropy spectroscopy (RAS) \cite{Aspnes1985_PRL, Martin2004_TSL, Bussetti2009_JVSTA} have been around for many years and offer a non-destructive way of probing such intermolecular interactions. In-situ DRS has been used to study electronic coupling and its evolution with thickness, for example, between flat-lying tin(II)-phthalocyanine (SnPc) molecules \cite{ Gruenewald2016_PRB} and between quaterrylene molecules \cite{Forker2012_APSC} deposited under vacuum. These works demonstrate the strength of in-situ optical spectroscopic techniques, but also one of the current limitations of DRS and RAS. While it is possible to infer electronic coupling in the perpendicular direction, for example, between quaterrylene molecules and the underlying gold layer (as a metal substrate) \cite{Forker2012_APSC}, due to the near-normal incidence of the probe light in DRS (and in RAS), the data obtained is primarily that of the in-plane component of the dielectric constants \cite{Forker2012_APSC} since the orientation of the electric field of the probe light is nearly all in the plane of the film. Being able to directly access the out-of-plane interactions of molecules would complement DRS/RAS and open up new avenues in probing and characterizing electronic coupling in interacting systems essential for further expanding the understanding of the physics of thin film optoelectronic devices. For example, in many organic electronic devices, key processes happen at planar interfaces parallel to the substrate, e.g. this is where the generation of free charge carriers happens in planar heterojunction organic solar cells.

One experimental technique which can probe this out-of-plane direction is in-situ spectroscopic ellipsometry (iSE), also sometimes referred to as "real-time" SE. This is possible firstly because of the high angle of incidence employed during measurements (typically 65$^{\circ}$). Secondly because of the use of polarized light, where the amplitude and phase change of the electric field upon interaction with the thin film are measured \cite{Fujiwara, Azzam1977}. The combination of these parameters carry more information than just the intensity measured in typical optical spectroscopic techniques, and which is also the most significant drawback of X-ray techniques where phase information is lost. Thus, SE is a thin film characterization technique typically used to characterize the primary optical excitations in terms of optical properties: dielectric constants or refractive indices which represent the quality of the thin films \cite{Fujiwara, Azzam1977}. In the case of organic solar cells, the technique is typically used to characterize the active layers for understanding structure-property relationships \cite{Hinrichs2012_TSL, Yokoyama2011_JMC,Klein2015_JPE,Schunemann2013_JPCC}, and for optical simulations to model absorption profiles within a device \cite{Burkhard2010_AdvMat}. 

iSE is an advanced application of SE which makes it possible to study the growth and evolution of the optical properties of multi-layered thin films, e.g. semiconductors deposited by thermal evaporation under vacuum or other vapor-deposition methods \cite{Yokoyama2010_JAP,Ye2020_JMCC, Heitz1998_TSL}. (i)SE, in general, has an angstrom-level sensitivity towards changes in thicknesses. The standard SE analysis procedure, or SSE in short, for obtaining optical properties from the iSE data relies on fitting a dielectric function model (using, for example, Gaussian and Lorentzian functions) to the data, and the typical analysis is initiated with the assumption that the film has the same properties along its thickness \cite{Fujiwara,Azzam1977}. However, robust and confident in-depth analysis such as variation of dielectric constants along the thickness direction is not possible because of lack of additional data (only one angle of incidence in iSE against different angles of incidence as is possible in ex-situ SE including at normal incidence). Thus, the information obtained from SSE analysis is determined  not only by the dielectric functions used during the fitting but also the type of model used to analyse single or multi-layered thin films. Moreover, from model fitting, even if any anomalies are seen in the dielectric constants especially along the growth direction due to interactions with other materials at the interfaces, the question of whether these anomalies are real or artifacts from fitting is posed. To ascertain the analysis, additional optical information such as about the in-plane dielectric constants obtained from in-situ DRS \cite{Forker2012_APSC, Gruenewald2016_PRB} is essential. 

In this work, we present an experimental methodology using iSE coupled with a new data analysis method for studying the evolution of the electronic wavefunctions pertaining to the optical excitations along the thickness or vertical direction as a function of thickness or, equivalently, number of molecules in vacuum-deposited thin films. By an empirical comparison to the optical properties (n + j k), we show that the new iSE data analysis method extracts information representative of the same for every layer (thickness resolution down to 1 \si{\angstrom} possible) deposited. The analysis does not involve any model fitting and is only based on tracking the differential change in the iSE data. Any changes in the electronic wavefunctions corresponding to the frontier orbitals under excitations due to overlap with that of the neighboring molecules - both same and different species – in the growth (thickness) direction is reflected in the information obtained from the analysis. We demonstrate the ability of the method by its application to the study of vacuum-deposited organic semiconductor thin films using fullerene (C60),  boron subphthalocyanine chloride (SubPc), boron sub-2,3-naphthalocyanine chloride (SubNc) and N,N,N',N'-Tetrakis-(4-methoxyphenyl)benzidine (MeO-TPD) on different underlying layers. 

We also evaluate and discuss the implications of these observations by complementing the analysis with a systematic approach for obtaining dielectric constants along the thickness direction using SSE data analysis, i.e. model fitting. Moreover, we demonstrate the angstrom-level sensitivity, simplicity and reproducibility of the method. To the best of our knowledge, this is the first time such significant modulations in the optical response due to electronic coupling have been observed and the method reported. Finally, we conclude with the general applicability of this method for studying other systems where electronic coupling through intermolecular interactions are crucial.

\section{EXPERIMENTAL SECTION}
The organic semiconductors SubPc (> 99\%, Lumtec), SubNc (> 99\%, Lumtec), C60 ($\sim$ 99.999\%, CreaPhys), molybdenum oxide (MoOx) (> 99.998\%, Lumtec) and MeO-TPD (> 99\%, Lumtec) were used as purchased. Substrates silicon wafer (with 23 nm SiO$_2$ on top, purchased from J.A.Woollam, for calibrating the ellipsometer) and ultrasmooth quartz substrate (WZWOPTICAG) were cleaned with soap and deionized water, and then sonicated in acetone and isopropanol (in this order) at 50 $^{\circ}$C for 20 minutes before drying with an air gun. Then the substrates were UV-ozone treated for 10 minutes and loaded into the vacuum chamber (CreaPhys) via a nitrogen glovebox. A Woollam RC2 spectroscopic ellipsometer was mounted at $\sim$ 65$^{\circ}$ angle of incidence onto the vacuum chamber. The thin films were deposited on the loaded substrates by thermal evaporation of the organic semiconductors (at rates 0.1-0.5 \si{\angstrom}/s). The in-situ measurements (210-1690 nm or 0.7-5.9 eV) were carried out with an acquisition time which ranged from 4-10 s or measurement of data for every $\sim$ 1-2 \si{\angstrom} increase in thickness.

Standard spectroscopic ellipsometry analysis (SSE) for deriving the refractive indices of each layer through model fitting to the data (and described in Section S2 in the supplementary information) was carried out in CompleteEASE software from J.A.Woollam company. Ex-situ SE data of some of the thin films after deposition was obtained, where required, at 55$^{\circ}$, 65$^{\circ}$ and 75$^{\circ}$ angles of incidence. The $\psi$ and $\Delta$ time series values measured through CompleteEASE were exported into text format and analysed in python using the equation described in the Results and Discussion section. 

\section{RESULTS AND DISCUSSION}
\subsection{DART}
In standard SE measurements of thin films, the data measured are the electric field amplitude ratio $\psi$ and phase difference $\Delta$ as a function of wavelength ($\lambda$) or energy ($E$) of the probe light defined as the ellipsometric ratio $\rho$ \cite{Fujiwara, Azzam1977}

\begin{equation}
\rho = tan\psi(E) e^{-j\Delta(E)}
\end{equation}

\noindent During iSE, $\rho$ is measured as a function of time \textit{t} or, equivalently, thickness \textit{d} in the case of monitoring the growth of vacuum-deposited thin films, which in our case are organic semiconductors. A first derivative of Equation 1 with respect to time (or \textit{d}) yields the following equation

\begin{equation}
\dfrac{\delta\rho}{\delta t} = e^{-j\Delta}[(sec\psi)^{2}\times\dfrac{\delta\psi}{\delta t} - j\times tan\psi\times\dfrac{\delta\Delta}{\delta t}]
\end{equation}

\noindent When the time interval between measurements is constant, which is typically the case in in-situ measurements, Equation 2 simplifies to 

\begin{equation}
\delta\rho_{t} = e^{-j\Delta_{t}}[(sec\psi_{t})^{2}\times\delta\psi_{t} - j\times tan\psi_{t}\times\delta\Delta_{t}]
\end{equation}

\noindent where
\begin{subequations}
\begin{align}
	& real(\delta\rho) = cos\Delta_{t}(sec\psi_{t})^{2}\delta\psi_{t} - sin\Delta_{t}tan\psi_{t}\delta\Delta_{t} \\
	& imaginary(\delta\rho) = -[sin\Delta_{t}(sec\psi_{t})^{2}\delta\psi_{t} + cos\Delta_{t}tan\psi_{t}\delta\Delta_{t}] 
\end{align}
\end{subequations}

\noindent The measured iSE data, $\psi$ and $\Delta$, can be numerically differentiated for $\delta\psi$ and $\delta\Delta$ yielding $\delta\rho$ at time \textit{t} or thickness \textit{d}. 

\begin{figure*}[t]
\includegraphics[scale=0.369]{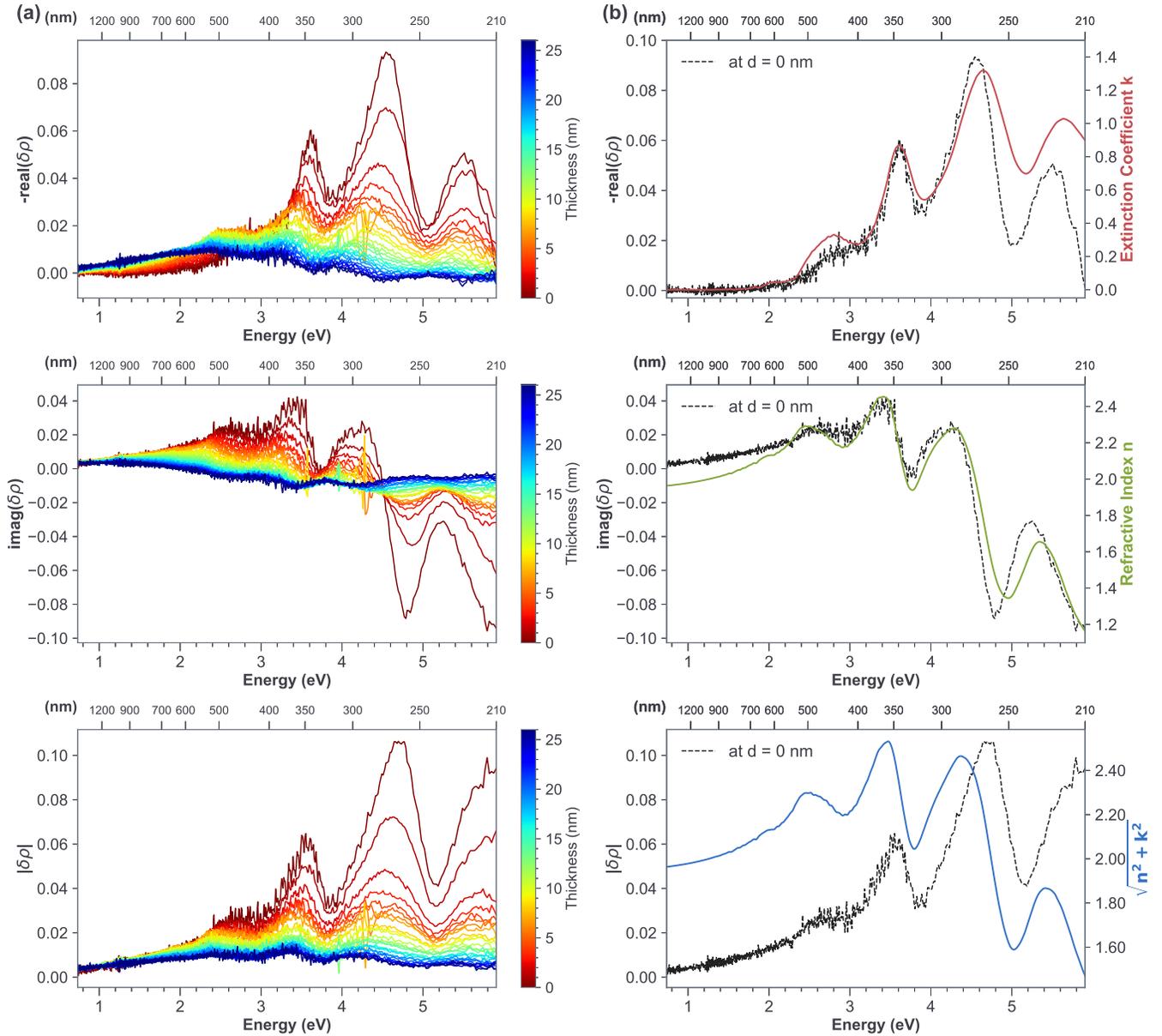}% Here is how to import EPS art
\caption{\label{fig:DART_C60_UQ} \textbf{(a)} DART analysis showing the real (negative) and imaginary parts as well as the modulus of $\delta\rho$ of C60 deposited on an optically smooth quartz substrate (SiO$_{2}$), calculated for a differential change of $\delta d$ = 1 nm in C60 thickness. \textbf{(b)} Comparison of the isotropic extinction coefficient k, refractive index n, and the modulus of n + j k of the same C60 film derived from SSE analysis at \textit{d} = 13 nm to the negative real, imaginary and modulus of $\delta\rho$ respectively of the 1st 1 nm layer deposited on the blank substrate (at “thickness” \textit{d} = 0 nm), taken from the DART analysis in (a).}
\end{figure*}

iSE is a highly sensitive technique, sensitive to changes in the thickness direction of a surface \cite{Fujiwara, Azzam1977} (also referred to as out-of-plane or z direction). In-plane/xy direction - probed in UV-visible absorption measurements - is defined as the plane of the film (see Figure S1). During iSE, the probe light passes through all the layers for every time point for which $\rho$ is measured. Since the thin film deposition or the direction of growth during iSE is along the z direction of the film, $\delta\rho$ then is the rate of change of the optical response $\rho$ of the thin film, which is primarily a function of the change in the optical properties of the thin film in the z direction, namely out-of-plane refractive indices N$_{z}$ (n$_{z}$+j k$_{z}$) or, equivalently, dielectric constants $\epsilon_{z}$ ($\epsilon_{1z}$ + j $\epsilon_{2z}$).  A basic derivation of $\delta\rho$ in terms of N$_{xy}$ (in-plane refractive indices of the film), N$_{z}$ and \textit{d} is given section S1 in the supplementary information. However, for a small, differential change in thickness $\delta d$, $\delta\rho$ can also be interpreted as the optical response of the incoming layer $\delta d$ that will be deposited on the substrate with the thin film at thickness \textit{d} as exemplified in Figures 1 and 2.  The following results are based on this method named as “Differential Analysis in Real Time” (DART).

\subsection{C60 and MeO-TPD on SiO$_{2}$}
\begin{figure*}[t]
\includegraphics[scale=0.369]{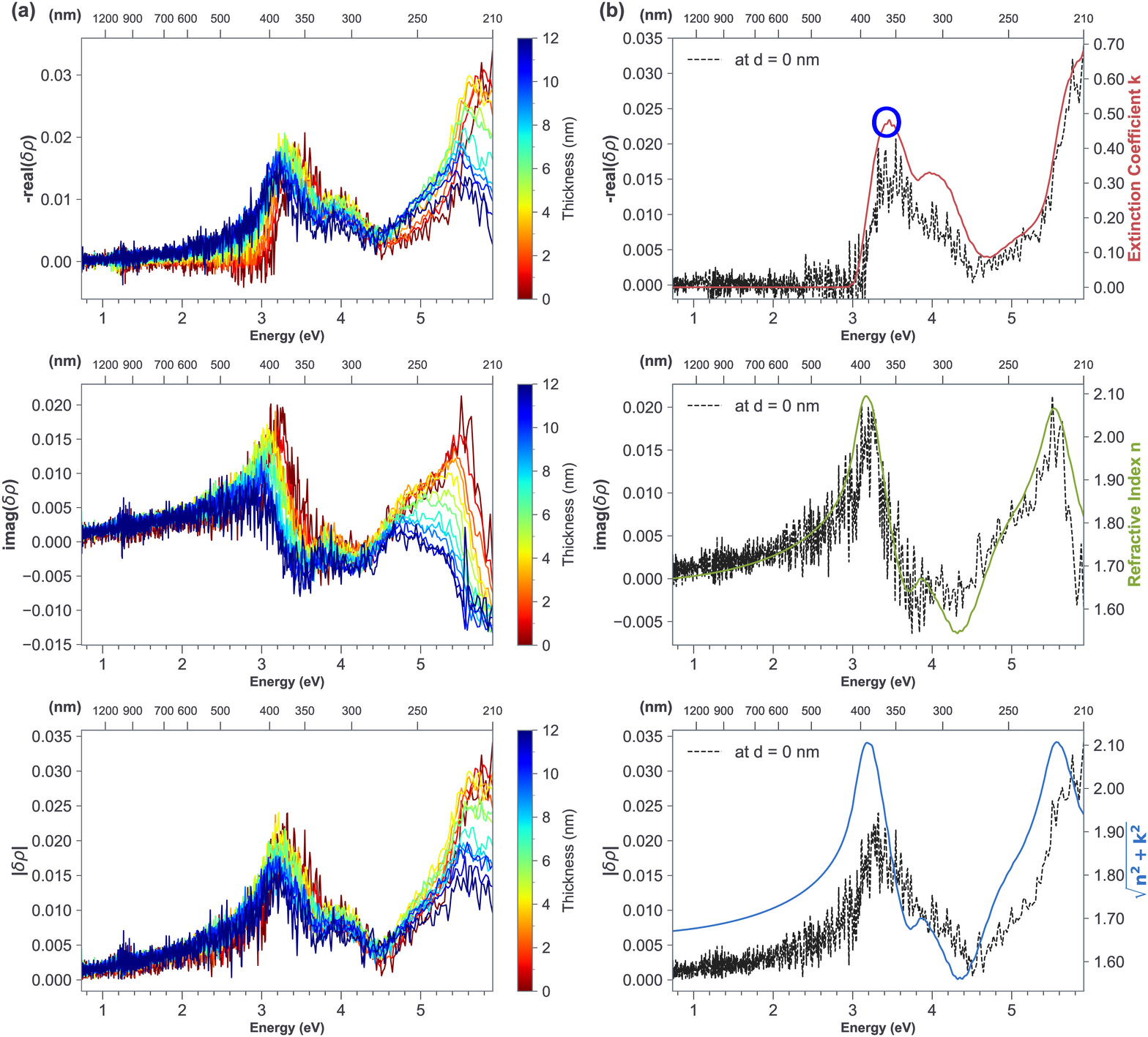}% Here is how to import EPS art
\caption{\label{fig:DART_MTPD_UQ} \textbf{(a)} DART analysis showing the real (negative) and imaginary parts as well as the modulus of $\delta\rho$ of MeO-TPD deposited on an optically smooth quartz substrate (SiO$_{2}$) calculated for a differential change of $\delta d$ = 1 nm. \textbf{(b)} Comparison of the isotropic extinction coefficient k, refractive index n, and the modulus of  n + j k of the same MeO-TPD film derived from SSE analysis at the final thickness of 12 nm to the –real($\delta\rho$), imaginary($\delta\rho$) and modulus($\delta\rho$) values respectively of the 1st 1 nm layer deposited on the blank substrate ("thickness" \textit{d} = 0 nm), taken from the DART analysis in (a). The $\pi-\pi$* transition at 3.4 eV is circled blue in the plot of extinction coefficient k in (b).}
\end{figure*}

Figure \ref{fig:DART_C60_UQ}(a) shows the real and imaginary parts as well as modulus of $\delta\rho$ of C60 deposited on an optically smooth quartz substrate (SiO$_{2}$); optically smooth implies that for the range of probe wavelengths (210-1690 nm), the roughness is on a sub-angstrom level. Thickness \textit{d} of the film is obtained from SSE analysis (see Section S2). $\delta\rho$ has been calculated for every 1 nm interval, i.e. $\delta d$ = 1 nm, for all the data presented in this work. Firstly, to understand the meaning of real and imaginary values of $\delta\rho$, we compare $\delta\rho$’s value calculated for the first $\delta d$ layer deposited on a blank substrate, i.e. at \textit{d} = 0 nm, to the k and n values of the C60 film (at 13 nm thickness) in Figure \ref{fig:DART_C60_UQ}(b) obtained from SSE analysis using Gaussian functions (commonly referred to as oscillators). Such a function/oscillator can be used to fit to the absorption peaks which represent the sum of the main electronic transition and vibrionic progressions. Examining the $\delta\rho$ profile of the first nanometer deposited, it can be seen that the -real($\delta\rho$) and imaginary($\delta\rho$) appear to represent k and n values respectively. The same observation can be seen for MeO-TPD (deposited on optically smooth quartz) in Figure \ref{fig:DART_MTPD_UQ}. Thus, $\delta\rho$ represents the optical properties of a layer $\delta d$, and its corresponding real (negative) and imaginary parts will be hereinafter referred to as $\delta\rho_{k}$ and $\delta\rho_{n}$ respectively.

SE probes the optical excitations of a system in terms of extinction coefficient k. Examining  $\delta\rho_{k}$ in Figure \ref{fig:DART_C60_UQ}(a) for the first 1 nm layer, which can be approximated as a monolayer of C60 molecules that is deposited on a blank substrate (\textit{d} = 0 nm), we observe that the strongest responses are centered at ~3.6 eV, 4.6 eV and 5.5 eV, which correspond to the allowed, primary transitions T$_{1u}$ $\leftarrow$ A$_{1g}$  with the strongest absorption in C60 \cite{Leach1992_ChemPhys,Sassara2001_Astro, Lucas1992_PRB, Pavlovich2010_JAS, Fukuda2012_JCP}; also reflected in the extinction coefficient plot in Figure \ref{fig:DART_C60_UQ}(b). This first nanometer data can be assumed as the average optical response of one C60 molecule deposited onto the SiO$_2$ substrate. As more C60 deposits, we observe that the optical response is not the same for every additional nanometer being deposited. The magnitude gradually decreases and the peaks red-shift. This can be attributed to the intermolecular interactions between C60 molecules in the form of electronic coupling arising from overlap (along the thickness direction) of the frontier orbitals corresponding to the respective primary transitions in C60. 

\begin{figure}[t]
\includegraphics[scale=0.43]{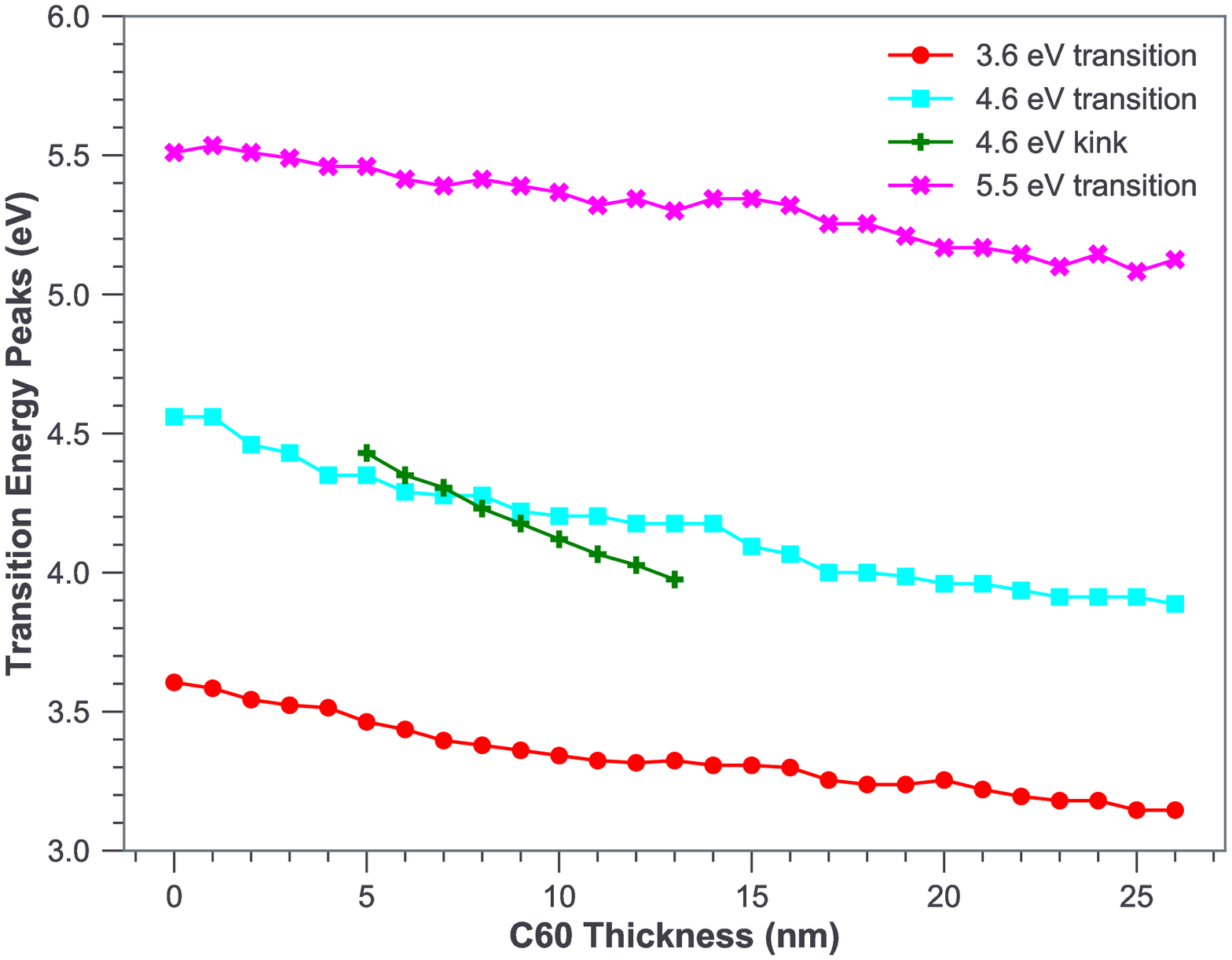}% Here is how to import EPS art
\caption{\label{fig:peaktrack_C60} Energy values of the C60 primary transition peaks as a function of C60 thickness obtained from the DART results in Figures 1 and S2.}
\end{figure}

A consequence of this coupling is the delocalization of the corresponding electronic wavefunctions under excitation (or excited state wavefunctions) in C60 -- a well-known phenomenon in C60 \cite{Bakulin2012_Science, Burlingame2018_Nature, Bernardo2014_NatComm,Kobayashi2020_JPCC} -- and analogous to band formation in inorganic semiconductors or J-type interaction \cite{Jelley1936_Nature, Scheibe1937_AChem} causing a gradual red-shift (along with polarization effects \cite{Kohler2015, Pope1999}) of the primary transitions with deposition. If there were no electronic coupling between the molecules, and consequently no delocalization, the response $\delta\rho$ should be the same for every deposition resulting in the same $\delta\rho$ for every $\delta d$. A constant $\delta\rho$ implies $\delta N_{z}$ = 0 (from Eq. S1.8) which is what is approximately observed for MeO-TPD for energies corresponding to the $\pi-\pi$* transitions at 3.4 eV in Figure \ref{fig:DART_MTPD_UQ}(a). The similar optical response at the $\pi-\pi$* transition energy for all thicknesses in MeO-TPD implies that the corresponding excited state wavefunction, unlike in C60, is highly localized indicating weak electronic coupling with neighbouring molecules at this energy. This is reflected in the extremely low mobilities of MeO-TPD \cite{Tietze2012_PRB,Meerheim2011_JAP}. However, the overlap of wavefunctions of the frontier orbitals in MeO-TPD occurs at higher energies as seen from the changes in $\delta\rho$ with deposition. These energies are higher than the ionization potential of MeO-TPD ($\sim$ 5.1 eV \cite{Olthof2012_APL}), and possibly correspond to transitions from energy levels below HOMO to LUMO and above.

\begin{figure*}[t]
\includegraphics[scale=0.39]{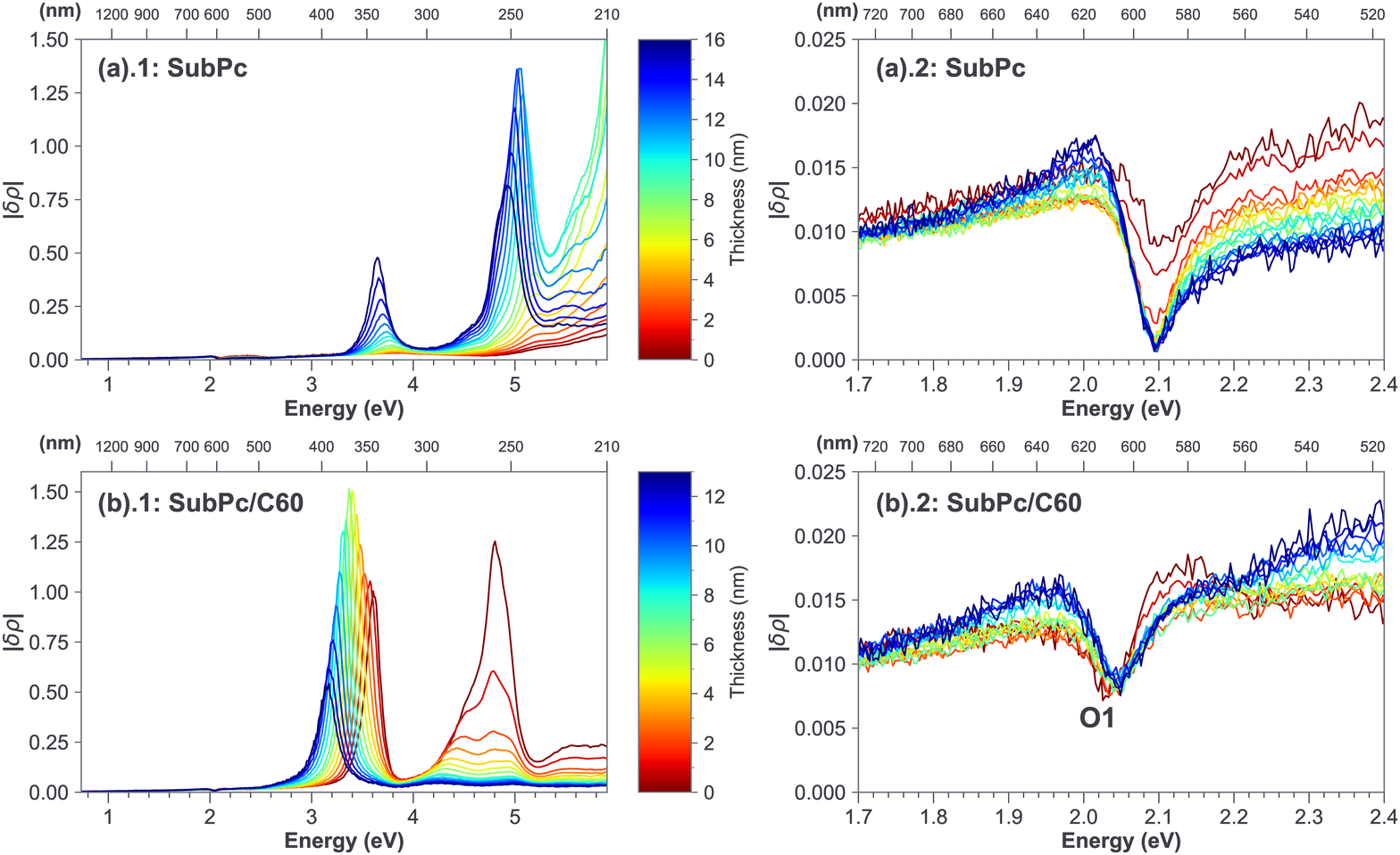}% Here is how to import EPS art
\caption{\label{fig:DART_SP_C60} \textbf{(a).1} DART analysis of a SubPc film deposited on a Si/SiO$_{2}$ substrate. \textbf{(a).2} Magnified view of the same DART results in (a).1 showing the optical response corresponding to the $\pi-\pi$* transition \cite{Morse2012_ACSAMnI, Rhoda2016_InChem} in SubPc for clear visualization. \textbf{(b).1} DART analysis of C60 deposited onto the SubPc film in (a). \textbf{(b).2} Magnified view of the data of (b).1 around 2.1 eV for clear visualization. The optical response corresponding to the excitations into SubPc excitonic states is indicated by the label O1, which corresponds to the transition in Figure S3(f). All calculations were carried out for a differential change of $\delta d$ = 1 nm. }
\end{figure*}

The dynamics of the three electronic transitions of C60 (Figure \ref{fig:DART_C60_UQ}(a)) seem to be different. Initially, the 4.6 eV transition has the highest strength. With addition of C60 (the film grows smoothly, see Section S2) the strength gradually decreases and goes below the 3.6 eV transition as shown in Figure S2, i.e. the rate of decrease of $\delta\rho$  of the 4.6 eV transition (also 5.5 eV) is higher compared to the 3.6 eV transition at least for the initial few nanometers; rate of change of $\delta\rho$ can itself be estimated from the second derivative of Equation 2 but not shown here because of the noisy output of the calculation. Additionally, a kink in the 4.6 eV transition appears at \textit{d} = 5 nm and persists until 13 nm or approximately a layer of 13 C60 molecules thick/high; the origin of this is not clear. It is possible that some sort of hybridization of the frontier orbitals corresponding to the 4.6 eV transition has occurred leading to its manifestation as a kink. Finally, analysis of the rates of red-shift of the transition peaks is shown in Figure \ref{fig:peaktrack_C60}. The 4.6 eV transition not only appears to decay faster (along with a kink) but also red-shifts faster (26 meV/nm) compared to the 3.6 eV (18 meV/nm) and the 5.5 eV (15 meV/nm) transitions. The broad, and weak, feature around 2.4 eV (below 3.0 eV), also red-shifting with addition of C60, is visible from the deposition of the first nanometer itself, and corresponds to a forbidden transition which occurs due to mixing of the vibronic states with the forbidden states \cite{Leach1992_ChemPhys,Sassara2001_Astro, Lucas1992_PRB, Pavlovich2010_JAS, Fukuda2012_JCP}. Such red-shifts of the transitions with thickness occurring due to coupling is a common feature in organic semiconductors as shown for other systems such as SnPc \cite{Gruenewald2016_PRB} and $\alpha$-quaterthiophene \cite{Bussetti2009_JVSTA}. Finally, In the energy range corresponding to the transparent region of C60 (from $\sim$ 750 nm or 1.7 eV), any such coupling should be absent, and is reflected in the similar and flat profile of $\delta\rho$. 

Such variation of $\delta\rho$ with thickness shows that the optoelectronic properties of C60 or any such interacting systems is not uniform along the z direction. Moreover, because the wavefunctions of C60 delocalize, the optical response $\delta\rho$ of a layer $\delta d$ deposited at time \textit{t} cannot be solely assigned to that layer $\delta d$ \cite{Bakulin2012_Science, Burlingame2018_Nature, Bernardo2014_NatComm,Kobayashi2020_JPCC}. The C60 molecules at the top and bottom of the film will be experiencing different environments, and thus coupling, compared to the molecule in the middle. Hence, the overall $\delta\rho$ could be the sum of all the changes in the underlying bulk layer (of thickness \textit{d}) along with the layer $\delta d$ deposited at time \textit{t}. This emphasizes that much care is needed when determining the optical constants along the thickness direction from SSE analysis thus highlighting its shortcomings; the values obtained are representative mainly of the properties along the xy or in-plane direction. Thus, the DART method can be used to track the optical properties of vacuum-deposited thin films in real time without the need for real-time model fitting of the iSE data which becomes highly complicated for multli-layered absorbing and anisotropic materials. However, the drawback is that the information obtained will be representative primarily of that of the optical properties in the out-of-plane direction. And to obtain the dielectric constants in interacting systems especially in the out-of-plane direction, as the results show iSE will be insufficient and will need to used in conjunction with other methods such as in-situ DRS \cite{Forker2012_APSC} to decouple in-plane and out-of-plane optical properties.

\subsection{SubPc}
DART analysis of a SubPc film deposited on a silicon wafer is shown in Figure \ref{fig:DART_SP_C60}(a) (only $\mid\delta\rho\mid$ is shown from here on as it represents n and k together). From the evolution of the data with thickness, we see that a strong electronic coupling between subsequent SubPc layers occurs at energies above 3 eV, i.e. between frontier orbitals lying much higher than for the 2.1 eV $\pi-\pi$* transition \cite{Morse2012_ACSAMnI, Rhoda2016_InChem}. Figure \ref{fig:DART_SP_C60}(a).2 is a magnified view of the data in Figure \ref{fig:DART_SP_C60}(a).1 showing the evolution of the optical response of the $\pi-\pi$* transition (the trough in $\mid\delta\rho\mid$) at 2.1 eV in SubPc. From the DART analyses of MeO-TPD and C60, a constant $\delta\rho$ with thickness implies negligible or weak delocalization. Thus, the electronic coupling between frontier orbitals corresponding to the $\pi-\pi$* transition  in SubPc appears to occur only until the 3rd or the 4th nanometer after which $\mid\delta\rho\mid$ at 2.1 eV remains constant indicating that the extent of delocalization of this excited state wavefunction is small, i.e. up to a maximum of 4 nm compared to at least 26 nm in C60 (Figure \ref{fig:DART_C60_UQ}). Corresponding SSE analysis of the completed SubPc thin film (16 nm) for the average in-plane and out-of-plane optical constants is shown in Figure S3 (e) and (f). As seen from this figure, an anisotropic model best describes the growth of SubPc (the mean squared error of the fit decreases from 5.4 for an isotropic to 2.1 for an anisotropic model), and a weak transition (O1) at $\sim$2 eV can also be seen.

\subsection{C60 on phthalocyanines}
Intermolecular interactions in the form of electronic coupling between C60 and phthalocyanines were next explored starting with SubPc as the bottom layer. C60 was evaporated onto the completed SubPc film (16 nm) of Figure \ref{fig:DART_SP_C60}(a) and iSE measurements were carried out during the deposition. The DART analysis is shown in Figure \ref{fig:DART_SP_C60}(b). Comparing to the growth of C60 on an insulating SiO$_2$ substrate (Figure \ref{fig:DART_C60_UQ}), the data shows strong, anomalous responses centered at 3.6 eV and 4.8 eV (with a shoulder at 4.6 eV). Moreover, with increasing deposition of C60, the response differs from that of C60 on SiO$_{2}$. The strength of the 3.6 eV transition increases with deposition of C60 until 6 nm, and remains strong, red-shifting and non-zero until the final thickness of 13 nm while that of the 4.8 eV decreases as in the case of C60 on SiO$_{2}$. Since $\delta\rho$ represents the optical response of the layer $\delta d$ that is deposited on the substrate with thin film at thickness \textit{d}, this implies that at C60 thickness \textit{d} = 0 nm on top of SubPc, the first incoming C60 molecule interacts (or electronically couples) strongly with SubPc at the 3.6 eV energy corresponding to the primary allowed electronic transition in C60. And as more C60 deposits, a strong overlap and delocalization of the excited state wavefunctions between the incoming C60 and C60 deposited on SubPc is occurring for the frontier orbital corresponding to the 3.6 eV transition. The red-shift of 3.6 eV is more prominent compared to C60 on SiO$_{2}$. DART analysis of the same data but calculated for a film thickness change of $\delta d$ = 1 \si{\angstrom} is shown in Figure S4 highlighting the angstrom-level sensitivity of the method.

Additionally, during deposition of C60 on SubPc, a $\mid\delta\rho\mid$ response can also be seen as a feature near 2.04-2.05 eV shown in the magnified view in Figure \ref{fig:DART_SP_C60}(b).2, which is not present in the C60 growth on SiO$_{2}$ in Figure \ref{fig:DART_C60_UQ} (and Figure S2). Thus, it appears that the origin of this feature is in the underlying SubPc layer. Comparison with the 2.1 eV transition in Figure \ref{fig:DART_SP_C60}(a).2 shows that this feature is red-shifted from 2.1 eV by about $\sim$ 60 meV, and appears to correspond to the O1 transition in Figure S3(f). The presence of this feature was also investigated by carrying out the DART analysis, calculated for $\delta t$ = 1 min, of the iSE data measured during the downtime between the end of SubPc deposition and start of the C60 deposition. The $\delta\rho$ values of the analysis are shown in Figure S5, and are essentially zero indicating that there was no other deposition on the SubPc film during this downtime. The weak, noisy features concentrated around 3.6 eV and 5 eV similar to the energies corresponding to the strong features of SubPc in Figure \ref{fig:DART_SP_C60}(a).1 suggests a possible mild rearrangement of SubPc molecules (probably on the surface). This observation also suggests that the DART analysis can be used to directly characterize the stability of a film. 

\begin{figure*}[t]
\includegraphics[scale=0.24]{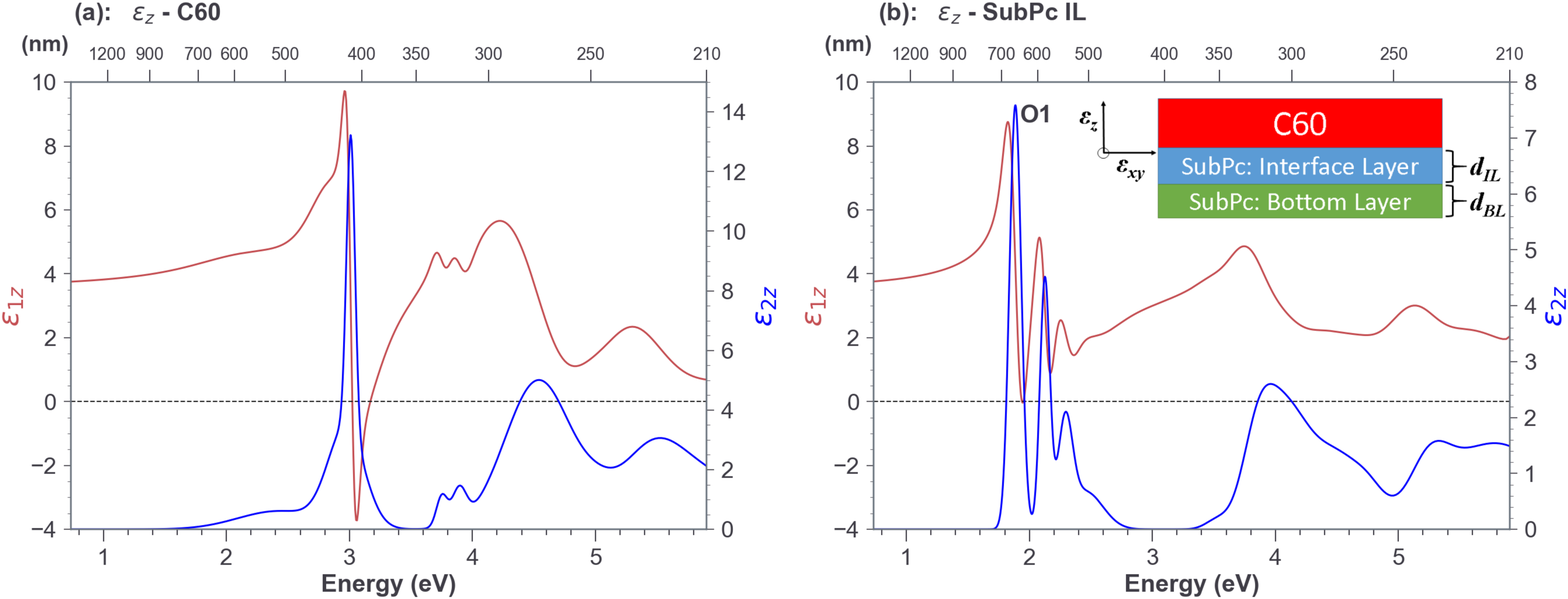}% Here is how to import EPS art
\caption{\label{fig:epsilon_SP_C60} The out-of-plane dielectric constants, $\epsilon_{z}$, of \textbf{(a)} 12 nm C60 layer on top of SubPc and \textbf{(b)} directly underlying (buried) 1 nm SubPc interface layer (IL$_{SubPc}$) showing the O1 transition at $\sim$ 2 eV. The layers correspond to that of Figure \ref{fig:DART_SP_C60}. Inset in (b) shows the schematic of the method used to derive the out-of-plane dielectric constants ($\epsilon_{z}$) of the interface layer of SubPc from standard spectroscopic ellipsometry data analysis method (SSE). The buried layer (SubPc here as an example) below C60  is divided into an interface layer with C60 of thickness $d_{IL}$ and a bottom layer $d_{BL}$. $\epsilon_{xy}$ is the in-plane dielectric constant of a layer. }
\end{figure*}

To gain an in-depth understanding of the significance of these anomalous optical responses, systematic analysis of the iSE data was carried out for obtaining the $\epsilon_{z}$ dielectric constants of C60 and the buried SubPc layers using the SSE method. For this, the buried SubPc layer was divided into an interface SubPc layer (IL$_{SubPc}$) and a bottom SubPc layer (BL$_{SubPc}$) as shown in the inset in Figure \ref{fig:epsilon_SP_C60}(b) (and Figure S6). Subsequently, the iSE data at 12 nm C60 thickness was fit for the out-of-plane dielectric constants $\epsilon_{z}$ ($\sqrt{\epsilon_{1z} + j \epsilon_{2z}}$ = n$_{z}$ + jk$_{z}$ of refractive index N$_{z}$) of the 12 nm C60 and the IL$_{SubPc}$ layer at thickness $d_{ILSubPc}$ = 1 nm shown in Figure \ref{fig:epsilon_SP_C60} (a) and (b) respectively; $d_{BLSubPc}$ was fixed at 15 nm with its $\epsilon_z$ set to the values before C60 deposition (Figure S3 (e) and (f)). $\epsilon_{xy}$ of the layers are fixed since the in-plane environment can be reasonably approximated as unvarying with thickness, i.e. $\delta \epsilon_{xy}/\delta d$ = 0 (see Section S2 for more details on the strategy). In contrast to the values of C60 on SiO$_{2}$ (Figure \ref{fig:DART_C60_UQ}) and SubPc on SiO$_{2}$ (Figure S3), the $\epsilon_{2z}$ of both C60 and IL$_{SubPc}$ here show significant increase at similar energies found in the DART analysis, i.e. around 3 eV and 2 eV respectively. More importantly, the corresponding real part $\epsilon_{1z}$ becomes negative at these values. This holds significant implications. In metal thin films such as Au, Ag and Al, and in conductive oxides such as indium tin oxide, the wavelengths where free electrons absorb are characterized by negative values of $\epsilon_{1}$ with high $\epsilon_{2}$ values due to absorption \cite{Fujiwara, Adachi2012_WorldScientific,Lynch1997}. The corresponding frequency of light where $\epsilon_{1z}$ equals to zero before taking on negative values is called plasma frequency. Hence, observation of such negative values of $\epsilon_{1z}$ and a concominant rise in $\epsilon_{2z}$ in both C60 and SubPc indicates that strong electronic coupling is occurring between the frontier orbitals corresponding to these specific energies - $\sim$3 eV in C60 and $\sim$2 eV in SubPc - and is leading to resonance electronic transitions in the form of plasmons (or unbound quasiparticles) oscillating between the molecules, and thus the delocalization of the excited state wavefunctions across the interface. A point to note is that the frontier orbitals corresponding to these excitations are at different energy levels relative to the vacuum level \cite{Olthof2012_APL,Cho2010_JPCC, Cnops2015_JACS}. The IL$_{SubPc}$ thickness at resonance will be referred to as $d_{ILSubPc}^{Res}$ from here on.

The parameters of the Gaussian oscillators (GO, given by amplitude: Amp, centre energy: En, broadening: Br) representing the two resonance transitions that were fit to the iSE data had good fit statistics, i.e. C60 GO -- Amp: 8.396$\pm$0.693, Br: 0.076$\pm$0.006, En: 2.997$\pm$0.003; IL$_{SubPc}$ GO -- Amp: 7.588$\pm$1.361, Br: 0.112$\pm$0.032, En: 1.886$\pm$0.024, with a low mean squared error (MSE $\sim$ 4.0) showing high confidence in the derived $\epsilon_{z}$ values. The value of $d_{ILSubPc}^{Res}$ (1 nm) obtained from SSE analysis, i.e. the depth in the SubPc film until which the resonating wavefunctions delocalize,  is lower than the 4 nm delocalization distance in SubPc obtained from the DART analysis in Figure \ref{fig:DART_SP_C60}(a).2. This is probably because of limitations of the SSE method requiring model fitting of a buried layer, and also lack of additional optical information to decouple the optical properties of the top C60 and the buried SubPc layers further highlighting the significance of the DART analysis method. However, $d_{ILSubPc}^{Res}$ can be taken as a lower estimate. Variation of $\epsilon_{z}$  of IL$_{SubPc}$ layer for different $d_{ILSubPc}$ values (1-16 nm) is shown in Figure S7, and similarly, variation of $\epsilon_{z}$ of the top C60 layer for its different thicknesses (9-13 nm) is shown in Figure S8, Table S1 for fit statistics of the parameters and Section S2 for more details.

In SubPc, this resonance absorption corresponds not to an increase in the $\pi-\pi$* transition but an increase in its adjacent O1 transition, which was originally observed in the $\epsilon_{2z}$ of the bulk film before C60 deposition (see Figure S3(f)). The O1 transition of SubPc at which the strong, plasmonic type coupling occurs with C60 corresponds to excitations into excitonic states. Its energy is below the $\pi-\pi$* transition in SubPc and in accordance with the energies of the SubPc excitons observed from fluorescence emission measurements \cite{Morse2012_ACSAMnI, Gommans2009_JPCC, Lin2014_SEMSC}. The observed 60 meV Stokes shift in our case, which is about 17 nm shift at 2.1 eV, is in agreement with the 8-30 nm (depending on the environment) observed in boron-subpthalocyanines \cite{Ye2020_JMCC, Morse2012_ACSAMnI, Kipp1998_JPCA, Morse2010_ACSAMI, Morse2011_ACSAMI, Helander2010_ACSAMI}. Since the excitons arise from the $\pi-\pi$* transitions -- also reflected in the decrease in $\epsilon_{2z}$ of the $\pi-\pi$* transitions at 2.1 eV and a concomitant increase in the O1 transition at resonance thickness $d_{ILSubPc}^{Res}$ in Figure S7 -- the electronic coupling, and its evolution, of C60 with the electronic wavefunction corresponding to the exciton suggests reorganization of SubPc molecules at the interface with C60. These observations are further corroborated by the DART results of C60 deposited on MoOx/SubPc (SubPc$_{MoOx}$) and MoOx/hexapropyltruxene (PrT)/SubPc (SubPc$_{PrT}$) shown in Figures S9 and S10 (SSE analysis for $\epsilon_{z}$ are shown in Figures S7 and S8). The Figures S9(b) and S10(b)  show the O1 transition at 2.04 eV and electronic coupling of C60 with SubPc similar to Figure \ref{fig:DART_SP_C60} further highlighting the robustness and strength of the DART analysis method. A major difference in the optical response of SubPc deposited on MoOx and PrT with that on SiO$_{2}$ can also be noticed at energies above 3 eV. We conjecture that this could be due to a possible electronic coupling between SubPc and the underlying electronic materials: MoOx and PrT respectively. 

In our previous work \cite{Ye2020_JMCC} we showed the use of PrT as an interface layer between MoOx and SubPc helping reduce exciton quenching leading to increase in the short-circuit current density ($J_{SC}$). Such interlayers are regularly used to improve the performance of organic solar cells \cite{Yin2016_AS,Bergemann2015_NL}. SSE analysis in that work also showed the O1 transition in $\epsilon_{2z}$ of both SubPc$_{MoOx}$ and SubPc$_{PrT}$ with that of the latter having the highest strength directly corroborating the increased number of excitons in SubPc due to PrT. This increase is also reflected in the increase in $\mid\delta\rho\mid$ value at 3.6 eV upon deposition of C60 on SubPc$_{PrT}$ (Figures S9(b).1 and S10(b).1). With higher number of excitons available in SubPc$_{PrT}$, higher number of resonance oscillations of the corresponding excited state wavefunctions are occurring with C60 leading to an increased $\mid\delta\rho\mid$ value. This correlation of increased number of excitons leading to an increase in the resonance $\mid\delta\rho\mid$ value is, to an extent, analogous to the observed increase in the absorption at the $\pi-\pi$* interchain transition from regiorandom P3HT to regioregular P3HT \cite{Brown2003_PRB,Spano2014_ARPC}. The red-shifts of the $\mid\delta\rho\mid$ peak energies of C60 as a function of C60 thickness are shown in Figure S11. The rates of shift are about the same while the peak values for C60/SubPc$_{PrT}$, C60/SubPc$_{MoOx}$, C60/SubPc$_{SiO2}$ (of Figure \ref{fig:DART_SP_C60}) differ by about 0.1 eV. However, compared to that of C60 on SiO$_{2}$ ($\sim$18 meV/nm), the rate of shift is about a factor of 2 higher: $\sim$40 meV/nm.

\begin{figure*}[t]
\includegraphics[scale=0.39]{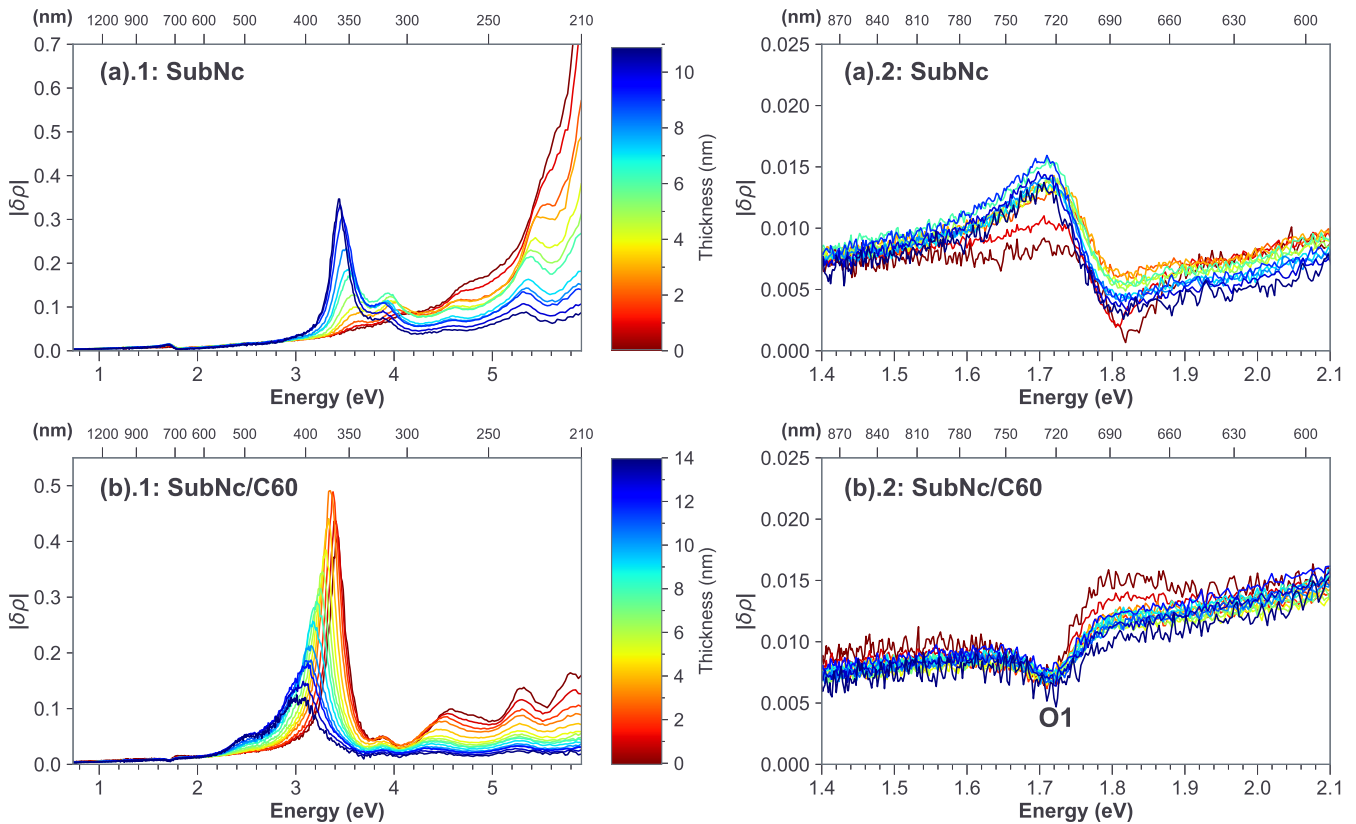}% Here is how to import EPS art
\caption{\label{fig:DART_SN_C60} \textbf{(a).1} DART analysis of a SubNc film deposited on Si/SiO$_2$/MoOx substrate. \textbf{(a).2} Magnified view of the same DART results showing the optical response corresponding to the $\pi-\pi$* transition in SubNc for clear visualization. \textbf{(b).1} DART analysis of C60 deposited on the SubNc film in (a). \textbf{(b).2} Magnified view of the same DART results in (b).1 around 1.8 eV. The optical response corresponding to the excitations into SubNc excitonic states is indicated by the label O1. All calculations were carried out for a differential change of $\delta d$ = 1 nm.  }
\end{figure*}

Finally, to ascertain that this type of electronic coupling phenomenon was not specific to SubPc and C60, experiments and data analyses were carried out for SubNc/C60 bilayer. Results similar to SubPc/C60 were observed, and shown in Figures \ref{fig:DART_SN_C60} (DART), S7 (SSE: $\epsilon_{z}$ of IL$_{SubNc}$ layer vs $d_{ILSubNc}$), S8 (SSE: $\epsilon_{z}$ vs C60 thickness), S11 (peak shifts) and Table S1 (fit statistics of the O1 transition). Figure \ref{fig:DART_SN_C60}(a) shows, as for SubPc, firstly, strong electronic coupling, and delocalization, occurring between frontier orbitals of SubNc (deposited on silicon wafer with 4 nm MoOx) at energies above 3 eV. For the wavefunction corresponding to the $\pi-\pi$* transition, the delocalization distance is about 2-3 nm after which $\mid\delta\rho\mid$ remains approximately constant. Figure \ref{fig:DART_SN_C60}(b) shows the same type of coupling as for SubPc/C60 between the wavefunction corresponding to SubNc exciton (O1) and that of the 3.6 eV primary transition in C60, along with delocalization, further corroborating the results. 

The plasmonic type absorption ($\epsilon_{2z}$) at very specific energies occurring upon deposition of C60 on the phthalocyanines implies resonance oscillations of the excited state wavefunctions between the two materials across the interface. The question of specificity in energies of the frontier orbitals at which electronic coupling is occurring can be partly investigated by examining the transitions in pure C60. The lower energy HOMO$\rightarrow$LUMO transitions are forbidden in C60 since the states are of the same parity. The primary electronic transitions are $^{1}T_{1u}$ $\leftarrow$ $^{1}A_{g}$ (or HOMO$\rightarrow$LUMO+1) and approximately centered around 3.6 eV, 4.6 eV and 5.5 eV \cite{Leach1992_ChemPhys, Lucas1992_PRB, Pavlovich2010_JAS,Fukuda2012_JCP, Orlandi2002_PPS, Montero2012_PCCP, Bauernschmitt1998_JACS}, and are also the wavefunctions which spatially delocalize between C60 molecules as seen in Figures \ref{fig:DART_C60_UQ}, \ref{fig:DART_SP_C60}, \ref{fig:DART_SN_C60}, S9 and S10. From our observations, it appears that the physics dictating the transitions within C60, i.e. Fermi’s golden rule \cite{Kohler2015, Pope1999, Dirac1927_PRSC}, also seems to dictate the transitions from other molecules -- arising from resonance oscillations -- into C60. More specifically, the question of why electronic coupling is occurring at 3.6 eV and not the other primary transitions is difficult to answer at this point. We can only conjecture which is that the frontier orbitals corresponding to the 3.6 eV primary transition is closest in energy, relative to the vacuum level, to that of the $\pi-\pi$* transition in the phthalocyanines. This coupling might have consequences for exciton dissociation from phthalocyanine to C60 and vice versa. In light of these intriguing observations, understanding their implications in the perspective of quantum mechanics and solar cell device physics is of pivotal importance, and will be explored in our next work with further donors and acceptors including non-fullerene acceptors, and at different substrate temperatures to explore the effects of molecular orientation on electronic coupling. The method will also be applied to test for Fermi-level pinning arising due to electronic coupling between organic semiconductors and metal electrodes.

\section{CONCLUSIONS}
We have shown a novel method for analysing in-situ spectroscopic ellipsometry data yielding direct information representative of the optical properties (dielectric constants) primarily along the growth direction of vacuum-deposited thin films in real time without the need for model fitting. Application of this analysis method to the study of pristine organic semiconductor thin films, C60 and MeO-TPD on SiO$_2$, showed electronic coupling and delocalization of the excited state wavefunctions corresponding to the frontier orbitals. Extending the same to the growth of C60 on phthalocyanines (SubPc and SubNc), strong, anomalous optical features were observed in C60 at energies corresponding to its primary allowed transition at 3.6 eV, which increased and then decreased in strength while red-shifting with C60 deposition. Accompanying features were observed in the underlying phthalocyanine layers at energies corresponding to their excitons. Dielectric constants along the growth or thickness direction of the bilayer films derived from systematic analysis using standard ellipsometry data analysis method (model fitting) revealed such features to be plasmonic type oscillations of the excited state wavefunctions corresponding to phthalocyanine excitons and the 3.6 eV primary transition of C60, the strength of which changed with the number of excitons. Spatial delocalization (along the thickness direction) of the wavefunctions in C60 and phthalocyanines was also observed. 

These results using the new analysis method have revealed singular insights into the physics of fullerene C60 and phthalocyanines in the thickness direction, and thus understanding the implications on organic optoelectronics device physics such as organic solar cells will be crucial. In the case of deposition techniques where in-situ growth measurements are not possible such as solution-processing methods, the DART method can still be used to study post-treatment effects, e.g. DART analysis could give  insights into effects of structural changes on the optical properties, for example, from thermal treatment, and consequently structure-property relationships and stability can be examined. And finally, strength of the presented analysis method as displayed in its simplicity, angstrom-level sensitivity, robustness and reproducibility, could be highly useful in conjunction with other in-situ optical spectroscopy methods \cite{Forker2012_APSC, Martin2004_TSL} in examining thin film systems comprising of semiconductors, quantum materials or any other materials where interfacial electronic coupling is highly crucial in determining the device properties.

\section*{SUPPLEMENTARY MATERIAL}
See the supplementary material for a basic derivation of $\delta\rho$; standard spectroscopic ellipsometry (SSE) analysis description and issues; derived dielectric constants; fit statistics of the O1 transition parameters; DART results of C60 on SiO$_{2}$, time gap between SubPc and C60, C60 on SiO$_2$/SubPc calculated for $\delta d$ = 1 \si{\angstrom}, C60 on MoOx/SubPc and PrT/SubPc; and $\mid\delta\rho\mid$ peak energy values of C60 vs C60 thickness deposited on the different SubPc films and SubNc film.

\section*{ACKNOWLEDGEMENTS}
SVK would like to thank the AFMD group, headed by MKR at Oxford, especially  Anna Jungbluth, Andreas E. Lauritzen, Dr. Alberto Privitera,  Irfan Habib, Dr. Pascal Kaienburg, Dr. Thomas Derrien for the valuable discussions and feedback on the manuscript. SVK and MKR would also like to thank Prof. Peter Skabara, School of Chemistry, University of Glasgow for hexapropyltruxene. SVK expresses gratitude to EPSRC (WAFT, Grant No. EP/M015173/1; IAA, Grant No. EP/R511742/1) and UKRI (START, Grant No. ST/R002754/1) for funding.   The research materials (iSE data) supporting this publication can be publicly accessed on the Oxford University Research Archive via the following digital object identifier: https://doi.org/10.5287/bodleian:KzjB5BnPb. The research materials are available under a CC BY license.

\section*{AUTHOR CONTRIBUTIONS}
SVK developed the DART method, and conceptualized and executed the study. Discussion of the results and planning the experiments were carried out by SVK and MKR. The first draft was written by SVK. All authors participated in the preparation and review.

\noindent \textbf{ORCID iD}\\
	SVK: 0000-0001-5526-0780 \\
	MKR: 0000-0002-5399-5510

\section*{COMPETING INTERESTS}
Oxford University Innovation has filed a UK patent application related to the DART method of this manuscript. Application No.: 2002576.3, Status: Pending.
  
%\bibliography{References_main}
\bibliographystyle{apsrev}

\end{document}